\documentclass[doublespacing]{elsart}

\usepackage{epsfig}

\begin{document}

\begin{frontmatter}



\title{\textbf{LAMBDA} \textbf{:} \textbf{L}arge \textbf{A}rea \textbf{M}odular \textbf{B}aF$_{2}$ \textbf{D}etector \textbf{A}rray for the measurement of high energy $\gamma$ rays}


\author[label1]{S. Mukhopadhyay},
\author[label2]{Srijit Bhattacharya},
\author[label1]{Deepak Pandit},
\author[label1]{A. Ray},
\author[label1]{Surajit Pal},
\author[label1]{K. Banerjee},
\author[label1]{S. Kundu},
\author[label1]{T. K. Rana},
\author[label1]{S. Bhattacharya},
\author[label1]{C. Bhattacharya},
\author[label3]{A. De},
\author[label1]{S. R. Banerjee\corauthref{cor}}
\corauth[cor]{Corresponding author.}
\ead{srb@veccal.ernet.in}


\address[label1]{Variable Energy Cyclotron Centre, 1/AF-Bidhannagar, Kolkata-700064, India}
\address[label2]{Dept. of Physics, Darjeeling Govt. College, Darjeeling-734101, India}
\address[label3]{Dept. of Physics, Raniganj Girls' College, Raniganj-713347, India}

\begin{abstract}
A large BaF$_{2}$ detector array along with its dedicated CAMAC electronics and VME based data acquisition system has been designed, constructed and installed successfully at VECC, Kolkata  for studying high energy $\gamma$ rays ($>$8 MeV). The array consists of 162 detector elements. The detectors were fabricated from bare barium fluoride crystals (each measuring 35 cm in length and having cross-sectional area of 3.5 $\times $ 3.5 cm$^{2}$). The basic properties of the detectors (energy resolution, time resolution, efficiency, uniformity, fast to slow ratio etc.) were studied exhaustively. Complete GEANT3 monte carlo simulations were performed to optimize the detector design and also to generate the response function. The detector system has been used successfully to measure high energy photons from $^{113}$Sb, formed by bombarding 145 and 160 MeV $^{20}$Ne beams on a $^{93}$Nb target. The measured experimental spectra are in good agreement with those from a modified version of the statistical model code CASCADE. In this paper, we present the complete description of this detector array along with its in-beam performance.

\end{abstract}

\end{frontmatter}

\section{Introduction}
The study of high energy gamma rays, coming from hot composite systems produced in heavy ion collisions is an important area in current nuclear physics research. High energy gamma photons originate mainly from the decay of giant dipole resonance (GDR) built on excited states and from nucleon-nucleon bremsstrahlung process in heavy-ion collisions. These photons are emitted at very early stages in the evolution or decay process of the hot  compound system and compete with the particle emission. These high energy photons can be used to study the properties of hot and fast rotating nuclei \cite{sno,gaar,van,status,fiss} and the collision dynamics in energetic heavy ion collisions \cite{nif}.

The high energy gamma rays interact with the detector material mainly through pair production and produce an electromagnetic shower. In order to confine this shower effectively, large volume scintillators with high efficiency for gamma detection are essential. Generally, NaI and BaF$_{2}$ scintillators are used for high energy gamma ray detection. Though energy resolution of NaI is better than BaF$_{2}$, they are comparable in detection efficiency. BaF$_{2}$ detectors have good timing property due to the presence of very fast scintillation decay component (600 ps). It is superior to NaI in all applications where fast timing is essential. It is also non hygroscopic and has a lower thermal neutron capture probability. During in-beam experiments, neutrons are a major source of contamination and they can only be discriminated by the time of flight technique. Detectors having better time resolution can be placed much closer to the target, thereby subtending larger solid angles (important for exclusive and low cross section measurements).

A high-energy gamma detector array is a very useful piece of research equipment. Most laboratories around the world have one form or the other \cite{karl,med,taps,barc}. At the Variable Energy Cyclotron Centre (VECC), Kolkata we have designed and developed a large BaF$_{2}$ detector system having 162 elements in a planar geometry along with remotely controlled dedicated CAMAC front-end electronics and VME based data acquisition system. A part of the array (49 detectors in a 7$\times$7 matrix) has been used for the measurement of high energy gamma photons from the reaction $^{20}$Ne + $^{93}$Nb at beam energies 145 and 160 MeV. In this paper, we report the complete description of the detector system and its in-beam performance.

\section{Array Design}
The detector array is composed of 162 square faced BaF$_{2}$ crystals. For the design of the array geometry, the following aspects were considered,
\vspace{.25cm}
\begin{enumerate}
\item Large gamma detection efficiency and fast timing response.
\item Granularity i.e., high segmentation to reduce pile up events.
\item Modularity, to allow different geometrical configurations depending upon the experimental requirements.
\item Easy maintenance and upgradation.
\item Compatibility with other detectors.
\item Cost-effectiveness i.e., to build from commercially available less expensive crystals.
\end{enumerate}
\vspace{.25cm}

\begin{figure}
\begin{center}
\epsfig{file=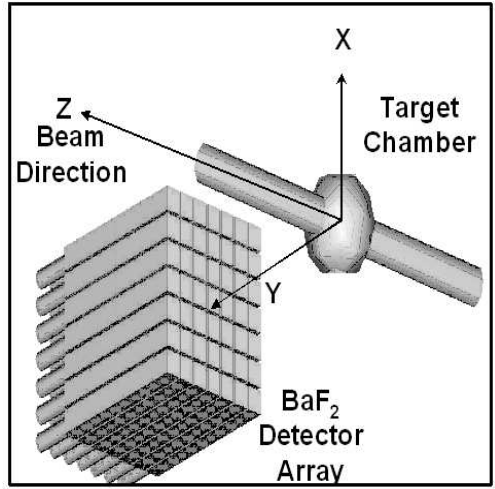,width=6.5cm,height=6.5cm}
\end{center}
\caption{Schematic view of detector array in 7$\times$7 matrix.}\label{setup}
\end{figure}

The higher granularity of this array greatly reduces the possibility of $\gamma$-$\gamma$ and $\gamma$-neutron pile up events. Being modular in nature, the detector elements can be arranged in two 9$\times$9, three 7$\times$7 or six 5$\times$5 matrix formations and placed at different angles with respect to the beam axis (important for GDR angular distribution studies as well as for coincidence experiments). Fig.[\ref{setup}] shows the schematic diagram of the detector set up for a 7$\times$7 matrix arrangement.

\section{Detector Fabrication}
Standard procedures were followed for detector fabrication from the bare barium fluoride crystals (each having dimension of 3.5 cm$\times$3.5 cm square faced and 35 cm long, procured from Beijing Glass \& Research Institute, China). The BaF$_{2}$ crystals have two scintillation light components ; a fast component ($\tau$=0.6 ns) peaking at $\lambda$=220 nm  and a slow component ($\tau$=630 ns) at $\lambda$=320 nm with intensities 20\% and 80\% respectively. The proper collection of the scintillation light is crucial to the time and energy resolution of the detectors. The crystals were cleaned properly using pure dehydrated ethyl alcohol and wrapped with 8-10 layers of 10 $\mu$m white Teflon cloth (C$_{2}$F$_{4}$) since it is a very good UV refelctor. Another advantage of using Teflon is that it acts as a diffused reflector, producing detector outputs nearly independent of scintillation centre position inside the crystal volume, minimizing the non-uniformity in case of detectors having larger lengths. Aluminium foil of thickness 10 $\mu$m (3-4 layers) was used to enhance the light collection and to block the surrounding light from entering into the detector. Fast, UV sensitive photomultipliers (29 mm dia, Phillips XP2978) were coupled with the crystals using a highly viscous UV transmitting optical grease (Basylone, $\eta $ -- 300000 cstokes). After coupling, the complete system was covered with black tape for light tightness. Specially designed aluminium collars of unique shape were used around the coupling area to provide additional support. Finally, for mechanical stability the whole assembly (crystal + PMT) was covered with heat shrinkable PVC tube. 
\par 
The pulse shape and the uniformity of the detectors were studied using a $^{137}$Cs standard $\gamma$-ray source and an oscilloscope (Tektronix model TDS724D, 500 MHz). The fast to slow component amplitude ratio, defined as maximum pulse amplitude divided by the amplitude of the signal measured at 20 ns from the peak position, was found to lie between 10 to 12 for all the crystals. The uniformity of the response was checked by moving a $^{137}$Cs source along the length of the crystals and the corresponding peak positions were monitored. A fraction of light was permitted to escape near the photomultiplier tube to achieve good uniformity. The non-uniformity was less than 5\% for all the detectors, but resulted in a slightly poorer resolution compared to uniform wrapping of the crystals.
\par
The $\alpha$-impurities present inside the crystals were checked using pulse shape discrimination technique and the measured values were less than 0.3 c/s/cc.

\section{ Calibration of individual elements in array during experiment}

\begin{figure}
\begin{center}
\epsfig{file=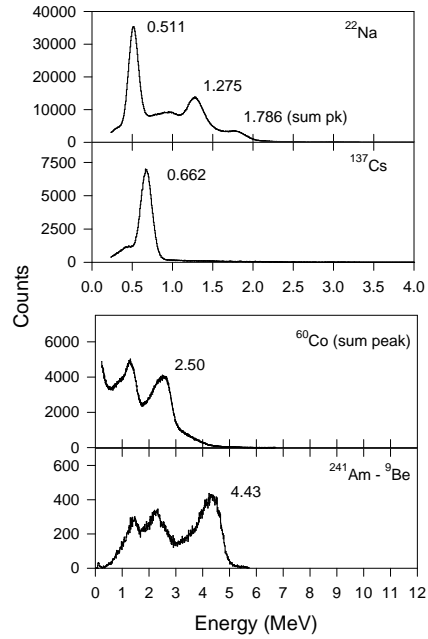,width=5.5cm}
\end{center}
\caption{Energy spectra from a single detector for different lab. standard gamma ray sources.}\label{spec}
\end{figure}

The detectors were calibrated using laboratory standard low energy gamma ray sources and 
minimum ionizing cosmic muons. The low energy points were obtained from $^{22}$Na (0.511 MeV, 1.274 MeV and sum peak 1.785 MeV), $^{137}$Cs (0.662 MeV), $^{60}$Co (sum peak 2.505 MeV) and $^{241}$Am-$^{9}$Be (4.43 MeV) sources. The gamma spectra measured in individual detectors are shown in Fig.[\ref{spec}].

\begin{figure}
\vspace{3.0cm}
\begin{center}
\epsfig{file=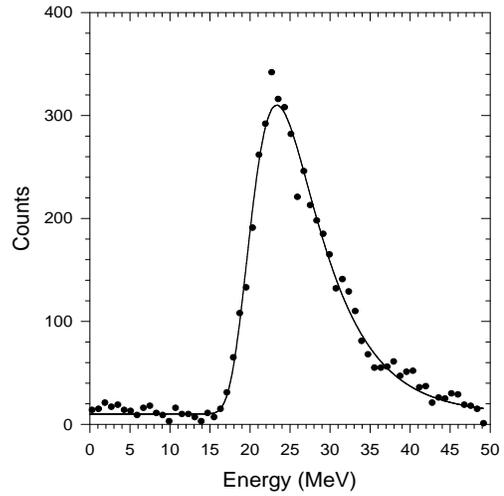,width=6.5cm,height=6.5cm}
\end{center}
\caption{Energy spectrum of the minimum ionizing cosmic ray muons in a single detector. The solid line is a fit through the data points using a Landau shape.}\label{cosmic}
\end{figure}

Minimum ionizing cosmic muons deposit 6.6 MeV/cm in BaF$_{2}$ material. While traversing the width of our detector (3.5 cm) vertically, they deposit a minimum of 23.1 MeV energy in the detector volume. Utilizing this fact, the whole array, arranged in a matrix form, can be calibrated in a very short time. A high threshold ($>$ 5 MeV) was set for all the detectors to ensure detection of only cosmic muons and the trigger was
generated only when all the seven detectors in a vertical column fired above this threshold. Independent spectra were recorded for all of the detectors showing identical response as expected. Following this procedure each detector of the array (in a 7$\times$7 matrix form) was calibrated in a run of four to five hours. Fig.[\ref{cosmic}] shows one such spectrum from a single detector. 

\par  

\begin{figure}
\vspace{1cm}
\begin{center}
\epsfig{file=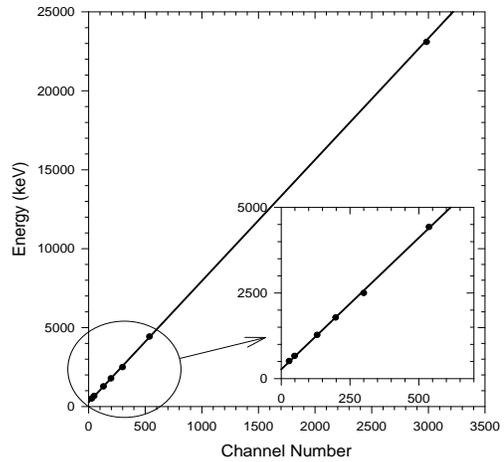,width=6.5cm,height=6.0cm}
\end{center}
\caption{Energy calibration curve for the BaF$_{2}$ detector with different standard $\gamma $ ray sources. Straight line fit to the low energy points (inset) is extrapolated up to 23.1 MeV.}\label{calib}
\end{figure}

The energy response of the detectors was found to be linear up to 4.43 MeV. The calibration curve is shown in Fig.[\ref{calib}] This calibration was extrapolated up to 23.1 MeV (energy lost by the cosmic muons) and found to match nicely.

\section{Energy and  time resolution of the individual elements}

\begin{figure}
\vspace{3cm}
\begin{center}
\epsfig{file=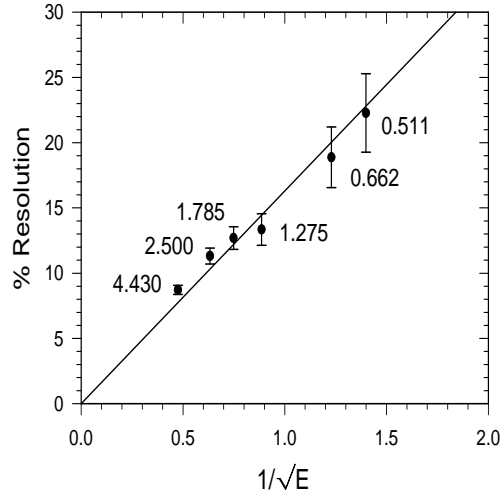,width=6.5cm,height=6.5cm}
\end{center}
\caption{Measured energy resolution plotted against 1/$\sqrt{E} $. The solid line shows the straight line fit passing through the origin.}\label{energy}
\end{figure}

The energy resolution of the detectors was calculated at different energies and plotted as a function of 1/$\sqrt{E} $ with its expected variation in Fig.[\ref{energy}]. A straight line passing through the origin is fitted (Res( \% ) = 16/$\sqrt{E} $) to the points.

\par

\begin{figure}
\vspace{2cm}
\begin{center}
\epsfig{file=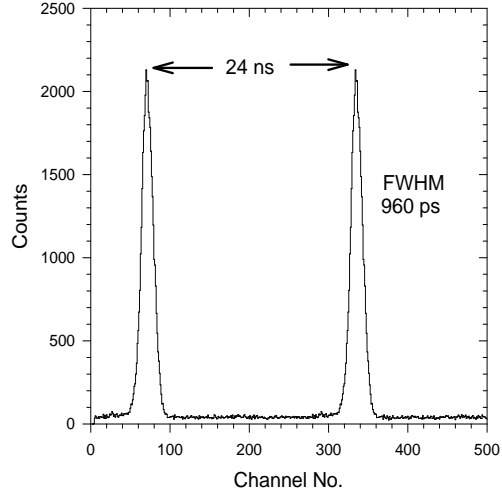,width=6.5cm,height=6.5cm}
\end{center}
\caption{The time calibration spectrum of one BaF$_{2}$ detector with respect to another similar one using $^{60}$ Co source for energy window 1 to 1.4 MeV.}\label{time}
\end{figure}

The time resolution between two BaF$_{2}$ detectors was measured with the $^{60}$Co source. The source was placed 
in between two identical detectors which were kept 180$^\circ$ apart. The energies and their relative times were measured 
simultaneously in event by event mode. The energy gated (1.0 to 1.4 MeV) time spectrum is shown in Fig.[\ref{time}]. The value obtained for time resolution is 960 ps.  

\section{Simulation studies for the detectors in array}

The response function of the detector array at different energies is necessary in order to fold a theoretically calculated gamma energy distribution for comparison with the experimentally measured high energy gamma spectra.
 In order to study the response of the array, simulations were done, considering the array in a realistic geometry, using the monte carlo code GEANT3 [13]. The detector energy resolution, experimental discriminator threshold values and wrapping thicknesses were incorporated properly into the simulation study. 

\par
\begin{figure}
\begin{center}
\epsfig{file=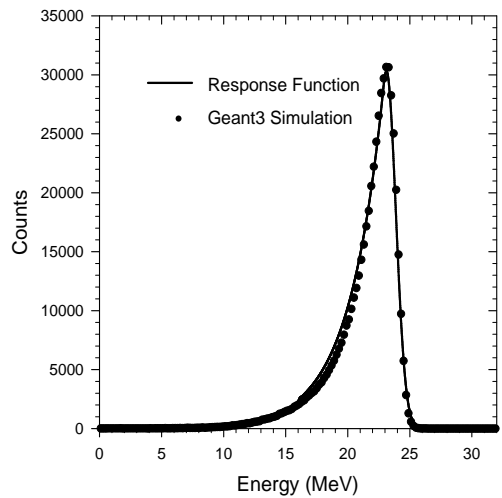,width=6.5cm,height=6.5cm}
\end{center}
\caption{The simulated line shape at E$_{\gamma}$= 24 MeV for the array in 7$\times$7 matrix compared with simulated response.}\label{res24}
\end{figure}
\begin{figure}
\vspace{3cm}
\begin{center}
\epsfig{file=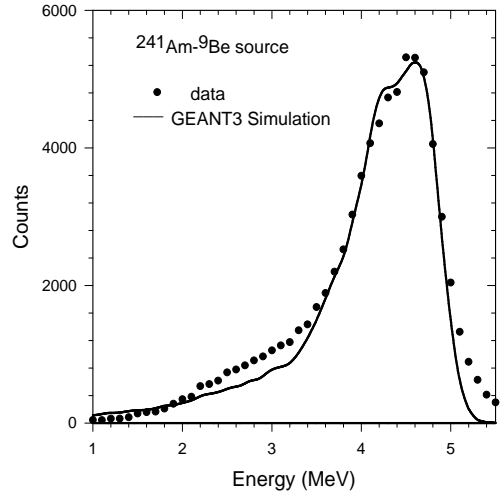,width=6.5cm,height=6.5cm}
\end{center}
\caption{Measured 4.43 MeV $\gamma$ energy spectrum along with simulated line shape.}\label{ambe}
\end{figure}

The array was configured in a 7$\times $7 matrix and was placed at a distance of 50 cm from the target. For event reconstruction different methods were tested and nearest neighbour summing or cluster summing was found to be the most suitable in this configuration. Firstly, the element was identified for maximum energy deposited above a high threshold ($>$ 3 MeV), and was checked for whether it had complete nearest neighbours. When both the above conditions were fulfilled, the energy deposits in the subsequent detectors in the cluster were added to reconstruct the event. The line shape of the array was studied for an energy range of 4 - 32 MeV. $\gamma$ rays with energies E$\gamma$ = 4, 6, 8, 10, 12, 16, 20, 24, 28, 32 MeV were generated originating from a point source at the target position. One hundred thousand events were thrown isotropically on the front face of the array to simulate the line shape of each $\gamma$ energy and one such spectrum at 24 MeV is  shown in fig[\ref{res24}]. The line shape at a particular energy was then parametrized using a gaussian and slope-matched exponential tail as shown in fig [7]. This procedure was repeated for all the given energies to find the energy dependent parameters. The energy dependence was then individually parametrized to generate the response function. The response function was tested with a measured spectrum of $^{241}$Am-$^9$Be source at 4.43 MeV. The simulated line shape matches well with the experimental one at 4.43 MeV (Fig.[\ref{ambe}])

\section{Electronics set up and DAQ}

A dedicated electronics setup has been developed to register the energy and time information from each detector element in event by event mode. To design this setup, it was decided to use the latest commercially available high-density CAMAC electronics. All electronics \& DAQ were kept inside the experimental hall next to the array and were controlled from outside over ethernet. 

An appropriate electronics scheme has been adopted for data collection. In this scheme, neutrons were eliminated by the time of flight technique and pulse shape discrimination   (long/short integration) was applied for pileup rejection in each detector element. It is well known that BaF$_{2}$ scintillation outputs are temperature sensitive \cite{temp}. Proper arrangements were made to record the ambient temperature around the detector system with an accuracy of less than 0.1$^\circ$C. In the VECC experimental hall the temperature was kept steady within 1$^\circ$C.

\par

Fixed gain 8-channel fast amplifiers (CAEN N412) were used to amplify the photomultiplier outputs. The amplified signals were split into two parts for the linear and the logic paths. The signals on the linear path were routed through a delay/attenuator/splitter box (designed and built in-house), where the analog signals were given a 100 ns delay each and were split into two parts with amplitude ratio (1:10) before sending to a pair of VME QDCs (CAEN V792) for long/short integrations. The long and short integration gates were selected as 2 $\mu$s and 50 ns respectively.
\par

\begin{figure}
\begin{center}
\epsfig{file=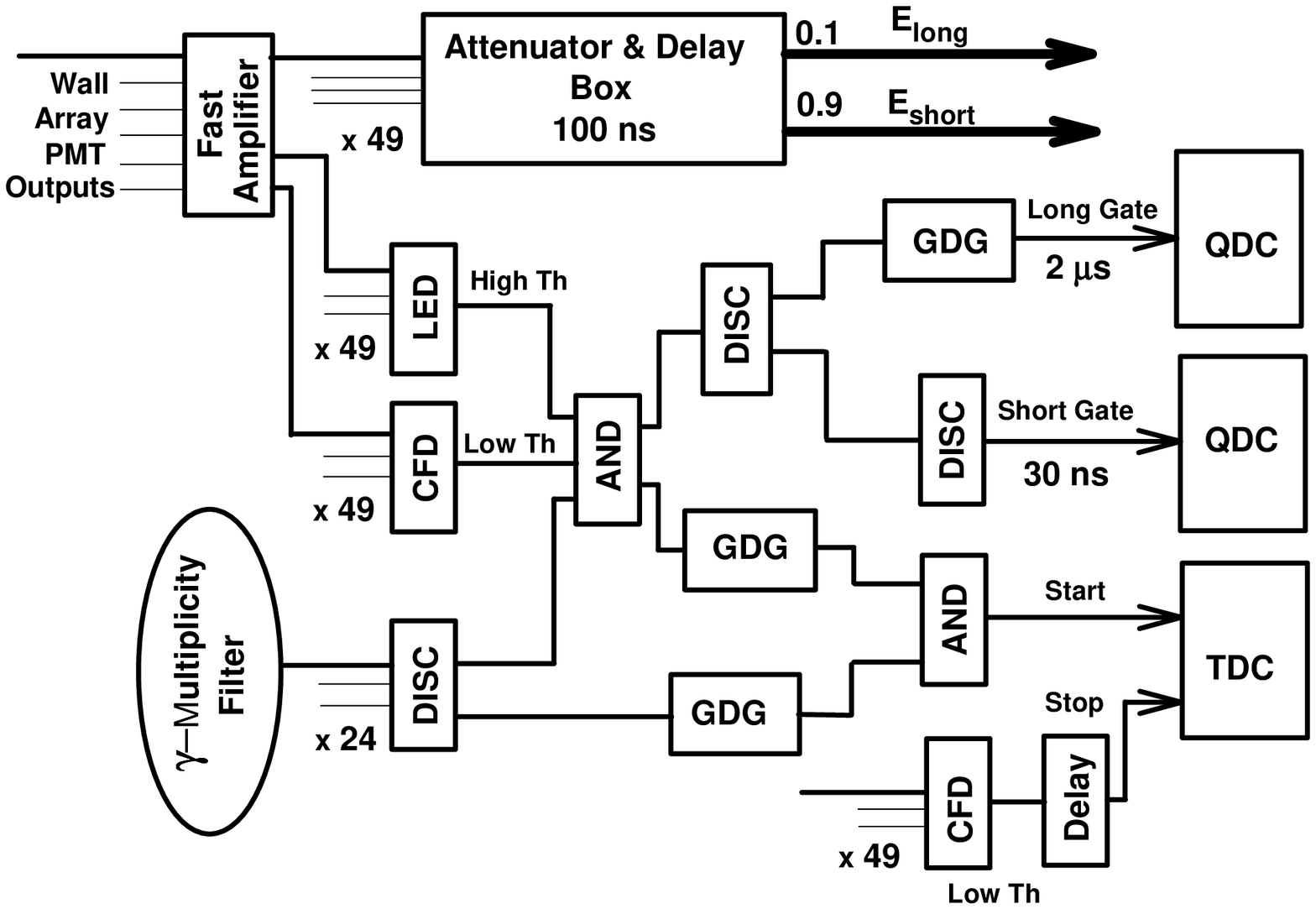,width=6.5cm,height=6.5cm}
\end{center}
\caption{The block diagram of the electronics set up}\label{circuit}
\end{figure}

For generating proper trigger conditions, the logic path signals were sent to CAMAC leading edge discriminators (LED) (CAEN C894) and constant fraction discriminators (CFD) (CAEN C671). The trigger was generated only when the fast component of the scintillation output crossed a certain high threshold T$_{h}$ set by the LEDs. The CFD outputs above a low threshold value T$_{l}$ were directly sent to the VME TDCs (CAEN V775) when T$_{h}$ was satisfied. The TDC common start/stop can be taken from either the RF signal of the Cyclotron or from the gamma multiplicity filter. The hit pattern in the detector array for high energy gamma rays and cosmic muons are different. The high energy photons produce a shower whereas cosmic muons  deposit a fixed amount of energy (6.6 MeV/cm) along its straight paths through the array. In triggered data acquisition mode (with starts from RF/multiplicity filter) the probability of cosmic events is small (rejection ratio $\sim$ 1:3300) and can be rejected effectively from the hit pattern by utilizing the square detector geometry and high segmentation of the array. In nearest neighbour summing mode, the cosmic muons will contribute to the higher energies, which is far above the energy range of our interest. The electronics circuit diagram is shown in Fig.[\ref{circuit}].

\par

The DAQ \cite{daq} associated with the array is VME controlled and works in a LINUX environment. High density, fast VME QDCs and TDCs have been employed for energy and time measurements. The data transfer rate was $\sim$4K/sec without appreciable dead time loss.

\section{In beam experiments}

\begin{figure}
\begin{center}
\epsfig{file=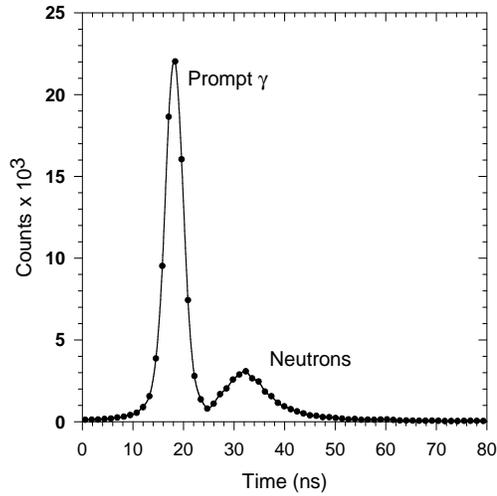,width=6.5cm,height=6.5cm}
\end{center}
\caption{The raw experimental time spectrum obtained from single detector.}\label{tof}
\end{figure}

\begin{figure}
\begin{center}
\epsfig{file=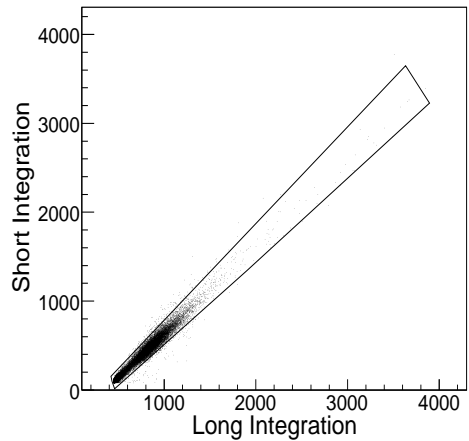,width=6.5cm,height=6.1cm}
\end{center}
\caption{Two dimensional long vs. short gate spectrum for a single detector for pulse shape discrimination along with appropriate cut.}\label{psd}
\end{figure}

An array of 49 BaF$_{2}$ detectors were arranged in a 7x7 matrix configuration as shown in Fig.[\ref{setup}]. This set-up was used for the measurement of high energy $\gamma$ spectra from the reaction $^{20}$Ne + $^{93}$Nb at beam energies 145 MeV and 160 MeV from K-130 AVF cyclotron at VECC. The detector pack was centered at 55$^\circ$ to the beam direction at a distance of 50 cm from the target. The array subtends a solid angle of 0.227 sr (1.8\% of 4$\pi$). Lead sheets of 3 mm thickness were placed in front and sides of the array, to cut down the intensity of the low energy gamma rays. The beam dump, located 3.5 metres downstream, was heavily shielded with borated paraffin blocks and lead bricks to cut down the gamma and neutron background. For the time of flight measurement, the time reference is taken from a 24 element gamma multiplicity filter (each having dimension of 3.5 cm $\times$ 3.5 cm  $\times$ 5 cm, BaF$_2$), split into two blocks of 12 detectors each and placed on two sides of the reaction chamber in a castle geometry for the estimation of angular momentum populated in each event. A clear separation between prompt gamma rays and neutrons was seen for all detectors and one such spectrum is shown in Fig.[\ref{tof}]. The trigger was generated by any one of the 49 detectors in the pack above a threshold of 4 MeV (to eliminate the background contribution) and the subsequent energy deposits were recorded for each detector above a low threshold of $\sim$250 keV. The pile up events were very few (Fig.[\ref{psd}]) due to the high granularity of the detector array and low event rate ($\sim$300/s) because of the exclusive trigger condition.
\par
\begin{figure}
\vspace{2cm}
\begin{center}
\epsfig{file=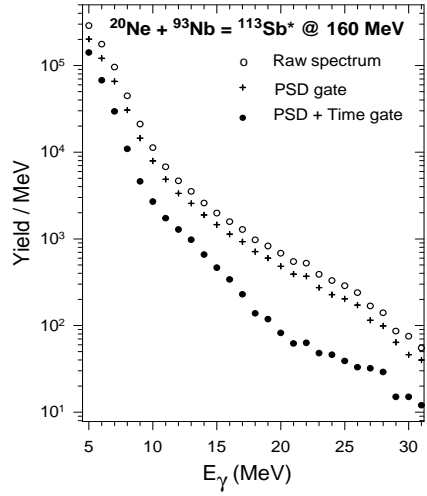,width=5.5cm,height=6.5cm}
\end{center}
\caption{The high energy gamma ray spectrum after applying PSD and prompt time cuts.}\label{psd_time}
\end{figure}
\begin{figure}
\vspace{2cm}
\begin{center}
\epsfig{file=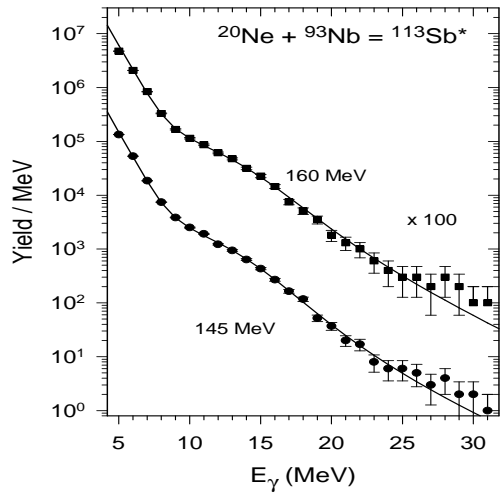,width=6.5cm,height=6.5cm}
\end{center}
\caption{Measured high energy gamma spectra at E$_{lab}$ =145 $\& $ 160 MeV along with statistical model (CASCADE) predictions.}\label{expt}
\end{figure}
\begin{figure}
\vspace{1.5cm}
\begin{center}
\epsfig{file=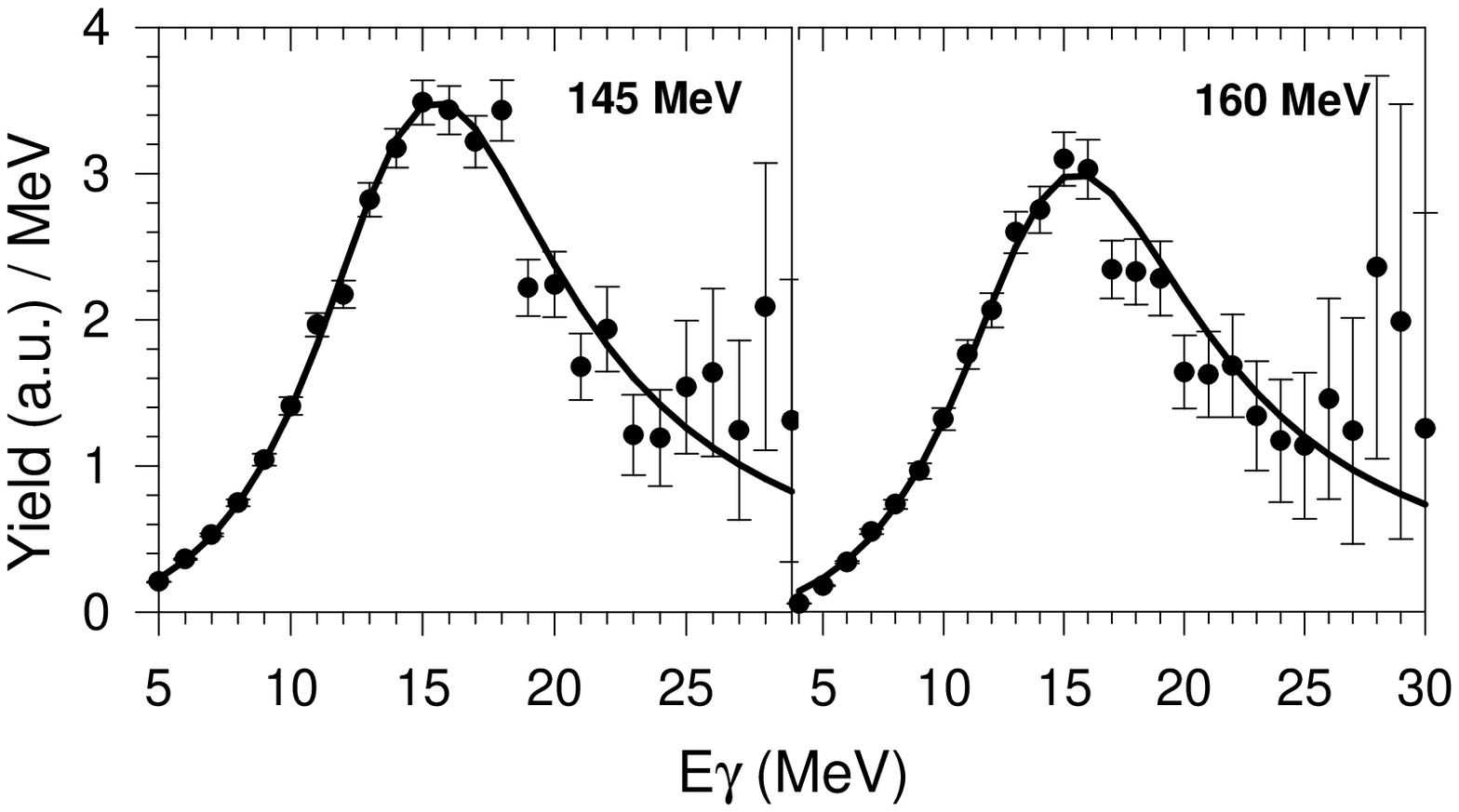,width=6.5cm,height=6.5cm}
\end{center}
\caption{Extracted linearized GDR plot along with statistical model (CASCADE) predictions.}\label{gdr}
\end{figure}

 The high energy gamma ray spectrum was reconstructed using the nearest neighbor (cluster) summing algorithm. Before summing all the nine elements of the cluster together, the energy deposits in each element were required to satisfy the gates/cuts employed viz., the two dimensional pulse shape discrimination (PSD) gate (Fig.[\ref{psd}]) and the prompt time gate. Fig.[\ref{psd_time}] shows the high energy $\gamma$-ray spectrum after applying the PSD and prompt time cuts. The experimental data (Fig.[\ref{expt}]) was compared with a modified version of statistical model code CASCADE \cite{cas} and was found to reproduce the data fairly well. The extracted linearized GDR plots are also shown in Fig.[\ref{gdr}].

\par

\section{Summary}
A highly granular and modular large BaF$_2$ detector array (162 elements) has been designed, fabricated and installed successfully for high energy gamma ray ($>$8 MeV) studies. The energy and timing characteristics of each detector element have been studied exhaustively. The GEANT3 monte carlo simulation has been performed considering realistic geometry to study the response of the array. The response function of the detector array in 7$\times$7 matrix formation is generated and tested for 4.43 MeV. The array has been used for the measurement of high energy gamma photon yields from the reaction $^{20}$Ne + $^{93}$Nb at beam energies 145 and 160 MeV. The experimental spectra are in good agreement with the statistical model code CASCADE.



\end{document}